\documentclass[12pt]{article}
\usepackage{graphicx}
\usepackage{subfigure}
\usepackage{cite}
\usepackage{amsmath}
\usepackage{capt-of}
\usepackage{array} 

\usepackage{tikz}
\usetikzlibrary{positioning,arrows.meta,shapes}

\begin{document}

\title{Ratio-Dependent Contrarian Activation in Opinion Dynamics}

\author{ Serge Galam\thanks{serge.galam@sciencespo.fr} \\
CEVIPOF - Centre for Political Research, Sciences Po and CNRS,\\
1, Place Saint Thomas d'Aquin, Paris 75007, France}

\date{(Entropy 2026, 28(4), 443)}

\maketitle

\begin{abstract}
I study the impact of mixed contrarians on the opinion dynamics of an heterogenous population with conformists using 
Galam Majority Model. Activation of contrarians is a function of the ratio majority/minority in the local groups of discussion. Restricting the group size to 3, two types of contrarians are included in respective proportions $c_{3,0}$ for configurations with ratio 3 to 0 and $c_{2,1}$ for  ratio 2 to 1. I then derive the explicit update Equation and obtained analytically the fixed points, their stability, and the resulting  full two-dimensional  landscape of the dynamics of opinion. Setting $c_{3,0} =c_{2,1} = c$ recovers the original results obtained with uniform contrarians.
The findings allow for considering a wide spectrum of new disruptive strategies to secure either a majority/minority ending ensuring the opinion having the larger initial support to win, or a single attractor dynamics at fifty/fifty, which implies a random winner regardless of initial supports.  

\end{abstract}


\section{Introduction}

Understanding the collective behavior of social systems is a central focus of sociophysics \cite{r1, r2, r3, r4, frank, book, bikas, inter, nun1, bik}. In particular, a great deal of works have been devoted to the study of opinion dynamics \cite{brazil, mak, p1, p2, p3, p4, p5, o1, o2, o3, o4, o5, o6, o7, nun2}. Early models investigated homogeneous groups of identical agents.  While each agent starts with their initial opinion, they all obey the same update rule of interactions and get their respective opinions evolve accordingly along update iteration. Among them stands the Galam Majority Model (GMM) introduced more than four decades ago \cite{min, tip}. The field keeps on being very active  \cite{b1, b2, b3, b4, b5, b6, b7, b8, b9, b10, b11, b12, b13, b14, b15, b16, b17}.

However, since real social systems are rarely fully homogeneous, subsequent models have introduced heterogeneity of behavior among agents when updating their respective opinions.

With this regard, I have introduced in 1997 the concept of biased agents  \cite{mosco}, latter defined equivalently as inflexible, stubborn, committed  and also zealots  \cite{stub, i1, i2, i3}. These agents do participate in the discussion groups with others like everyone else, but at the end they do not update their opinion according to the local majorities. They do stick to their initial opinion never shifting from it. Agents who obey majority rules are denoted equally as rationals, floaters and conformists. In the following, I use the term conformist.

Along another path of heterogeneity, in 2004 I have introduced contrarians in opinion dynamics \cite{contra, chaos, fake, asym}. Contrarians oppose the local majority rule contrary to conformists who adopt it.  They are activated according to some probability, which is independent of the initial local configuration of the discussion group. Thus, contrarians generate local fluctuations with respect to the deterministic majority rules used for updating individual opinions when agents are discussing in a small group. Many works have investigated the role of contrarians in opinion dynamics \cite{c1, c2, c3, c4, c5, c6, c7, c8, c10, c11, c12, c13}.

In this work, I extend the notion of contrarian by making its activation dependent on the initial majority/minority ratio of opinions in the local discussion groups. The study is restricted to discussion groups of size 3 with two competing opinions A and B. Contrarian behavior is symmetrical with respect to A and B.

The size 3 yields 4 different configurations of opinions. By symmetry only two different ratios are possible, unanimity and two against one. Therefore, two types of contrarians are introduced alongside  conformists. 

With three agents sharing initially the same opinion, either A or B, a contrarian is activated with a probability $c_{3,0}$ to oppose the majority. The agent behaves as a conformist with  probability $(1-c_{3,0})$. However, for two agents holding the same opinion against one agent holding the opposite opinion, a contrarian is activated with probability $c_{2,1}$ to oppose the majority. The agent behaves as a conformist with  probability $(1-c_{2,1})$. The parameters $c_{3,0}$ and $c_{2,1}$ are independent of each other.

Uniform contrarians have a linear impact of the dynamics with one critical proportion at which the landscape of opinion is turned up side down from a tipping point type dynamics into a single attractor dynamics. Here, mixed contrarians generate a two dimensional area in which a similar impact is achieved. The finding opens novel flexibility to implement the change of the dynamics landscape with a larger spectrum of strategies to monitor contrarians.

In particular, new disruptive strategies can be designed to secure majority avoiding a single attractor dynamics at fifty/fifty. Setting $c c_{3,0} =c_{2,1} = c$ recovers the original results obtained with uniform contrarians, i. e., contrarians being activated independently of the ratio of opinions.

The rest of the paper is organized as follows: The Galam Majority Model of opinion dynamics and the original subsequent inclusion of contrarians are reviewed in Sections 2 and 3. The investigation of two types of contrarians as a function of the local majority/minority ratio is done in Section 4. Last Section highlights the opposite available strategies to implement as a function of both the level of contrarian behavior and the initial majority and minority supports.

\section{The Galam Majority Model of opinion dynamics}

The Galam majority model of opinion dynamics (GMM) illustrates how local interactions can amplify small meaningless biases and lead to the phenomenon of democratic minority spreading \cite{min, tip}. Despite its simplicity, the model captures essential nonlinear mechanisms. 

In the case of homogeneous populations where agents (conformists) obey local majority rule to update their respective opinions while discussing in small groups, the dynamics are found to exhibit democratic tipping-point behavior with the tipping-point located at fifty percent. 

While application of local majority rule to small discussion groups eliminates minorities in favor of the initial aggregated majority, a tie-breaking prejudice in even-sized groups drastically disrupts the democratic balance. As a result, the original tipping point at fifty percent splits into two asymmetric tipping points, located above and below fifty percent, with the lower one driving the opinion dynamics that benefit from the prejudice \cite{min}. However, in the present paper, even groups do not appear.

The simplest nontrivial case of the GMM considers discussion groups of size 3, in which each agent holds one of two opinions, A or B with respective initial proportions $p_0 \in [0,1]$ and $(1-p_0)$. The dynamics are implemented by update iterations of individual  opinions according to the following scheme:
\begin{enumerate}
 \item Randomly dividing the population into groups of size 3.
 \item Applying a majority rule inside each group.
 \item All agents in the group adopt the opinion of the local majority.
 \item All agents are reshuffled.
\end{enumerate}

Four configurations are then possible as shown in Table (\ref{t3}) with their respective probabilities of occurrence. Thus, the associated update Equation for one sequence writes,
\begin{eqnarray}
\label{p3} 
p_{n+1} &=& p_n^3 + 3p_n^2(1-p_n)  ,\nonumber \\ 
&=& -2p_n^3+3p_n^2  ,
\end{eqnarray}
where $p_{n+1}$ is the new proportion of agents holding opinion A after n consecutive updates with starting proportion $p_0$. The associated fixed points driving the dynamics are $p_B = 0$, $p_c = \frac{1}{2}$, and $p_A = 1$, where $p_c$ is a tipping point and $p_{A,B}$, i. e.,  $p_A$ and $p_B$, are attractors. 

\begin{table}[h]
\begin{center}
\begin{tabular}{| c | c | c |}
\hline
Configuration & Probability & Majority \\
\hline
AAA & \(p_n^3\) & A \\
AAB, ABA, BAA & \(3p_n^2(1-p_n)\) & A \\
ABB, BAB, BBA  & \(3p_n(1-p_n)^2\) & B \\
BBB & \((1-p_n)^3\) & B \\
\hline
\end{tabular}
\caption{The four possible configurations of agents holding opinion A or B for a group of size three with a proportion $p_n$ for opinion A.}
\label{t3}
\end{center}
\end{table}


\section{Adding contrarians to the GMM}

In 2004 I have extended the GMM by introducing contrarians \cite{contra}, agents who systematically adopt the opinion opposite to
the local majority in their local group. Accordingly, if the majority in a group is A, conformists adopt or keep A but
contrarians adopt B. If the majority is B conformists adopt or keep B but contrarians adopt A.

Denoting $c$ the proportion of contrarians, the proportion of conformists is $(1-c)$. Contrarians are assumed to be randomly distributed and to become active only after the group majority has been determined. Accordingly, Eq. (\ref{p3}) is rescaled as,
\begin{equation}
p_{n+1} = (1-c)\left[p_n^3 + 3p_n^2(1-p_n)\right] + c \left[(1-p_n)^3 + 3p_n(1-p_n)^2\right]   ,
\label{pc33}
\end{equation}
where the first term accounts for the contribution from conformists (local A majorities) and the second for the contribution from contrarians (local B majorities). Eq. (\ref{pc33}) reduces to, 
\begin{equation}
p_{n+1} = (1-2c)\left[p_n^3 + 3p_n^2(1-p_n)\right] + c   .
\label{pc3}
\end{equation}

The presence of contrarians weakens the impact of majority rule by preventing the existence of unanimities obtained at  $p_B=0$ (all agents hold opinion B) and $p_A=1$ (all agents hold opinion A). The associated dynamics are identified by solving the new fixed point Equation $p_{n+1}=p_n$ using Eq. (\ref{pc3}) instead of Eq. (\ref{p3}).

By symmetry the fixed point $p_c = \frac{1}{2}$ is recovered but $p_{A,B}$ are modified with,
\begin{equation}
p_{B,A}= \frac{(1-2c) \pm \sqrt{12c^2-8c+1} } {2(1-2c)}  ,
\label{r3} 
\end{equation}
instead of (0, 1), and exist only in the range $ 0\leq c\leq \frac{1}{6}$ \cite{contra}. At $ c=\frac{1}{6}$, $p_A$ and $p_B$ coalesce with $p_c$ turning it to the unique fixed point of the dynamics in the range $ \frac{1}{6} < c \leq 1$. There, a recent work unveiled a rich behavior of the dynamics \cite{thwar}.

In the range $\frac{1}{6} < c \leq \frac{1}{2}$, $p_c$ is the unique attractor of the dynamics. For $\frac{1}{2} < c \leq \frac{5}{6}$ is still the unique attractor but associated now with an alternating regime. However, for $\frac{5}{6}<c\leq 1$, $p_c=\frac{1}{2}$ turns back to a tipping point although without attractors. The explanation of this apparent paradox was shown to be the setting of two alternating attractors \cite{thwar}. These findings have been revealed studying the stability of $p_c$ by expanding $p_{n+1}$ around $p_c$ using Eq. (\ref{pc3}), which yields,
\begin{equation}
p_{n+1} \approx p_c+ d_3 (p_n-p_c)+\dots
\label{dp1} 
\end{equation}
with, 
\begin{eqnarray}
\label{dc3} 
d_3 &=& \frac{d p_{n+1}}{d p_n}\Big|_{p_c}  ,\nonumber \\ 
&=& \frac{3}{2}(1-2 c) .
\end{eqnarray} 
Then, rewriting Eq. (\ref{dp1}) as $(p_{n+1}-p_c) \approx  d_3 (p_n-p_c)+\dots$ and iterating it down to $p_0$ gives, 
\begin{equation}
(p_{n+1}-p_c ) \approx d_3^{n+1} (p_0-p_c)+\dots .
\label{dp2} 
\end{equation}
which shows that $p_c$ is an attractor when $-1< d_3<1$ and a tipping point when $d_3<-1$ or $d_3>1$. These conditions are respectively satisfied in the range $\frac{1}{6} < c < \frac{5}{6} $, $c > \frac{5}{6}$ and $c< \frac{1}{6}$. For $c=\frac{1}{6}$ and $c = \frac{5}{6}$, $p_c$ is a unique fixed point and is an attractor.

In the case $c > \frac{5}{6}$ ($d_3<-1$), solving the Equation $p_{n+1}=1-p_n$ yields $p_c = \frac{1}{2}$ and two alternating fixed points,
\begin{equation}
p_{\bar{B},\bar{A}}= \frac{(1-2c) \pm \sqrt{12c^2-16c+5} } {2(1-2c)}  ,
\label{rbar3} 
\end{equation}
which are valid in the range $c \geq \frac{5}{6}$ in agreement with above result \cite{thwar}.

\section{Extension to two different types of  contrarians}

In this work, I make the activation of a contrarian behavior depend on the local ration majority/minority in the discussion groups. In the case of discussion groups of size 3, preserving the symmetry between A and B, two types of contrarians are thus introduced alongside  conformists. 

Contrary to the original behavior of  contrarians, here I make the activation of a contrarian to depend on the local ratio of majority/minority. A contrarian is now sensitive to the initial configuration of the discussion group and not to the final unanimity produced by the application of majority rule.

When all three agents share initially the same opinion, either A or B, a contrarian is activated with a probability $c_{3,0}$ to oppose the majority. The agent behaves as a conformist with  probability $(1-c_{3,0})$. When the initial ratio in the group is two agents holding the same opinion against one agent holding the opposite opinion, a contrarian is activated with probability $c_{2,1}$ to oppose the majority. The agent behaves as a conformist with  probability $(1-c_{2,1})$. Contrarian activations are identical for A and B majorities. 

To build the associated update Equation I first consider configurations AAA (probability $p_n^3$), which give AAA after application of majority rule. Then, contrarian behavior is activated with either zero contrarian, one contrarian, two contrarians, or three contrarians, with respective probabilities $(1-c_{3,0})^3, \; 3 c_{3,0} (1-c_{3,0})^2, \; 3 c_{3,0}^2 (1-c_{3,0}), \; c_{3,0}^3$ as shown in Table (\ref{ta3}).

\begin{table}[h]
\renewcommand{\arraystretch}{1.6}  
\setlength{\extrarowheight}{0.1cm} 
\hspace{-2cm}
\begin{tabular}{|c|c|c|c|c|}
\hline
Initial & Contras & Transition & Probability & Contribution (A, B) \\ [0.2cm]
\hline 
AAA & 0 & $ \to AAA$ 
& $(1-c_{3,0})^3$ 
& $ (1-c_{3,0})^3,\; 0 $ \\ [0.2cm]
\hline
AAA & 1 & $ \to BAA, ABA, AAB$ 
& $3c_{3,0}(1-c_{3,0})^2$ 
& $  \frac{2}{3}\cdot 3 c_{3,0}(1-c_{3,0})^2,\;  \frac{1}{3}\cdot 3 c_{3,0}(1-c_{3,0})^2 $ \\ [0.2cm]
\hline
AAA & 2 & $\to BBA, BAB, ABB$ 
& $3c_{3,0}^2(1-c_{3,0})$ 
& $ \frac{1}{3}\cdot 3 c_{3,0}^2(1-c_{3,0}),\; \frac{2}{3}\cdot 3 c_{3,0}^2(1-c_{3,0}) $ \\ [0.2cm]
\hline
AAA & 3 & $\to BBB$ 
& $c_{3,0}^3$ 
& $ 0,\; c_{3,0}^3 $ \\ [0.2cm]
\hline
\end{tabular}
\caption{Detailed contributions to opinions A and B starting from configuration AAA under contrarian behavior with respectively zero, one, two, three contrarians (contras).}
\label{ta3}
\end{table}
Adding all the various contributions to A from Table (\ref{ta3}) yields, 
\begin{equation}
(1-c_{3,0})^3 + 2c_{3,0}(1-c_{3,0})^2 + c_{3,0}^2(1-c_{3,0}) = (1-c_{3,0}),
\label{ca3}
\end{equation}
making $(1-c_{3,0}p_n)^3$ the total contribution to A from configuration AAA, while the complementary contribution to B is $c_{3,0} p_n^3$.

Table (\ref{tb3}) shows the equivalent output when starting from configurations BBB (probability $(1-p_n)^3$). Adding all the various contributions to A fromTable (\ref{tb3}) yields, 
\begin{equation}
c_{3,0}(1-c_{3,0})^2 + 2 c_{3,0}^2(1-c_{3,0}) + c_{3,0}^3= c_{3,0} ,
\label{cb3}
\end{equation}
making $c_{3,0} (1-p_n)^3$ the total contribution to A from configuration BBB while the complementary contribution to B is $(1-c_{3,0})(1-p_n)^3$.

\begin{table}[h]
\renewcommand{\arraystretch}{1.6}  
\setlength{\extrarowheight}{0.1cm} 
\hspace{-2cm}
\begin{tabular}{|c|c|c|c|c|}
\hline
Initial & Contras & Transition & Probability & Contribution (A, B) \\ [0.2cm]
\hline 
BBB & 0 & $ \to BBB$ 
& $(1-c_{3,0})^3$ 
& $ 0, \; (1-c_{3,0})^3 $ \\ [0.2cm]
\hline
BBB & 1 & $ \to ABB, BAB, BBA$ 
& $3c_{3,0}(1-c_{3,0})^2$ 
& $  \frac{1}{3}\cdot 3 c_{3,0}(1-c_{3,0})^2,\;  \frac{2}{3}\cdot 3 c_{3,0}(1-c_{3,0})^2 $ \\ [0.2cm]
\hline
BBB & 2 & $\to AAB, ABA, BAA$ 
& $3c_{3,0}^2(1-c_{3,0})$ 
& $ \frac{2}{3}\cdot 3 c_{3,0}^2(1-c_{3,0}),\; \frac{1}{3}\cdot 3 c_{3,0}^2(1-c_{3,0}) $ \\ [0.2cm]
\hline
BBB & 3 & $\to AAA$ 
& $c_{3,0}^3$ 
& $ c_{3,0}^3, \; 0 $ \\ [0.2cm]
\hline
\end{tabular}
\caption{Detailed contributions to opinions A and B starting from configuration BBB under contrarian behavior with respectively zero, one, two, three contrarians (contras)}.
\label{tb3}
\end{table}

By symmetry  with AAA and BBB the total contributions to A from AAB, ABA, BAA and ABB, BAB, BBA, are $(1-c_{2,1})3p^2(1-p)$ and $c_{2,1}3p(1-p)^2$ where $c_{2,1}$ replaces $c_{3,0}$, $3p^2(1-p)$ replaces $p^3$, and $3p(1-p)^2$ $(1-p)^3$.

From above results, adding the various contributions to A turns the update Equation (\ref{pc3}) to,
\begin{eqnarray}
\label{pcc3} 
p_{n+1} &=& (1-c_{3,0})  p_n^3 + (1-c_{2,1}) 3 p_n^2 (1-p_n) + c_{2,1} 3 p_n (1-p_n)^2  + c_{3,0} (1-p_n)^3  , \nonumber \\ 
&=& (1 + c_{3,0}- 3 c_{2,1} ) (-2p_n^3+3p_n^2)- 3 ( c_{3,0}-c_{2,1}) p_n +c_{3,0}  ,
\end{eqnarray}
where the effect of heterogeneous contrarians on the Equation can be seen by comparing with  Eq.  (\ref{pc3}). In the first term $ (1 + c_{3,0}- 3 c_{2,1} )$ replaced $(1-2c)$, a term in $p_n$ has appeared and $c_{3,0}$ replaced $c$. Setting $c_{3,0}=c_{2,1}=c$ recovers Eq.  (\ref{pc3}).

\subsection{Fixed Points and their domain of existence}

\begin{figure}[h]
\centering
\includegraphics[width=1\textwidth]{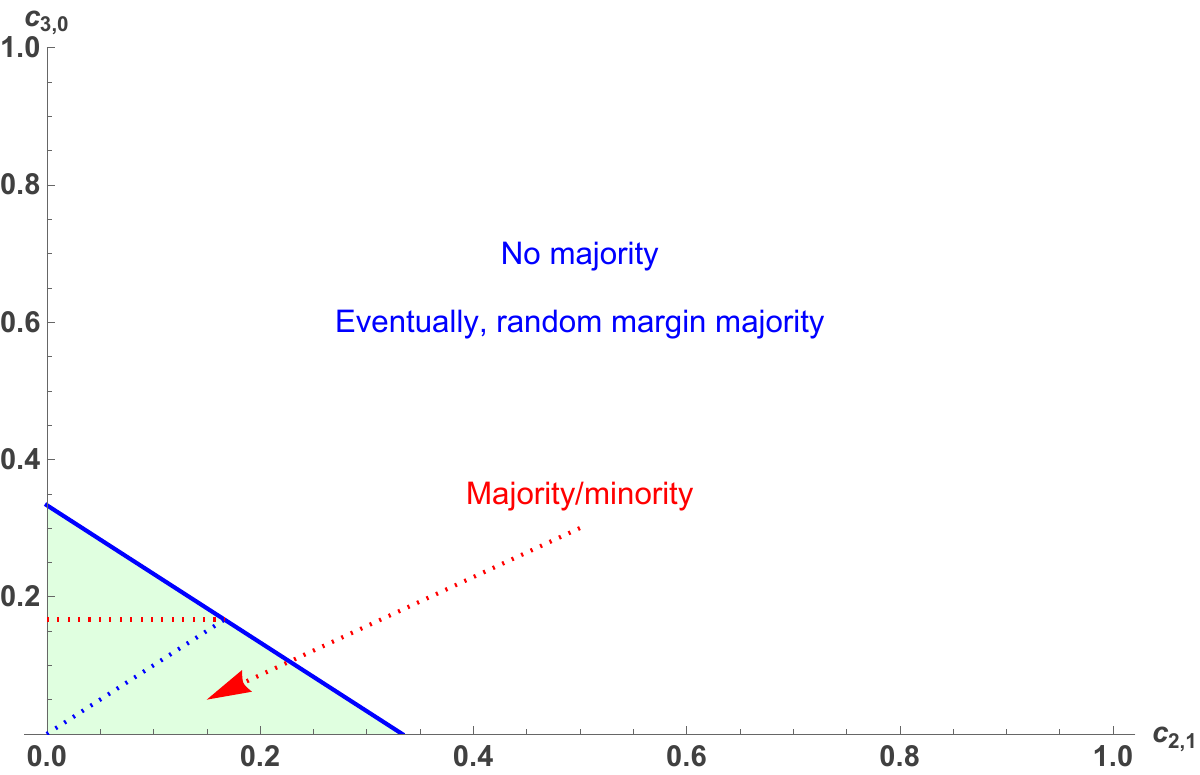}
\caption{The two domains of the landscape of the dynamics induced by the variation of  $c_{3,0}$ and $c_{2,1}$. The domains are separated by the line $c_{3,0} = \frac{1}{3} (1 - 3  c_{2,1})$. Inside the colored area, $p_{A,B}$ exist and are attractors with $p_c$ being a tipping point. Outside the colored area in the white area only the fixed point $p_c=\frac{1}{2}$ exists and is the single attractor of the dynamics.}
\label{r0}
\end{figure}

Using Eq. (\ref{pcc3}), the fixed point Equation $p_{n+1}=p_n$ yields again $p_c=\frac{1}{2}$ by symmetry but the two fixed points $p_{A,B}$ from Eq. (\ref{r3}) are modified with,
\begin{equation}
p_{A,B}=\frac{1}{2} \Biggr[ 1 \pm \sqrt{\frac{-1 + 3  c_{2,1} + 3  c_{3,0}}{-1 + 3  c_{2,1} -  c_{3,0}}} \Biggl] ,
\label{rr3} 
\end{equation}
which exist only in the combined range $ c_{2,1} \leq \frac{1}{3}$ and $0 \leq c_{3,0} \leq \frac{1}{3} (1 - 3  c_{2,1})$. Two different qualitative domains are thus found and shown in Figure\ref{r0}. 

The domains are separated by the line $c_{3,0} = \frac{1}{3} (1 - 3  c_{2,1})$. Inside the colored area, $p_{A,B}$ exist and are attractors with $p_c$ being a tipping point. There, the dynamics of opinion end up with one opinion being majority against the other one being minority. Initial $p_0 >\frac{1}{2}$ leads to A victory with an ending support $p_A>\frac{1}{2}$. On the contrary, $p_0 < \frac{1}{2}$ leads to A being defeated with a final support $p_B<\frac{1}{2}$.

Along the line $c_{3,0} = \frac{1}{3} (1 - 3  c_{2,1})$, both attractors $p_{A,B}$ merge with $p_c=\frac{1}{2}$, which turns from tipping point to a single attractor. Beyond this line, in the white area the dynamics are driven by the single attractor $p_c$ making any initial supports for respectively A and B, ending at precisely equal proportions of fifty percent. There is no majority, the population being perfectly polarized with the coexistence of two balanced opposite supports. However, it is worth to stress that the polarization is not frozen but fluid with agents keeping shifting opinion but in equal proportions \cite{pola}. Moreover, in case of a vote, any meaningless mistake in ballot counting, even lower than the statistical error, would lead to a margin random victory of either A or B. The two domains are exhibited in Figure\ref{r0} \cite{hung}.

\begin{figure}[h]
\vspace{-3cm}
\centering
\includegraphics[width=0.48\textwidth]{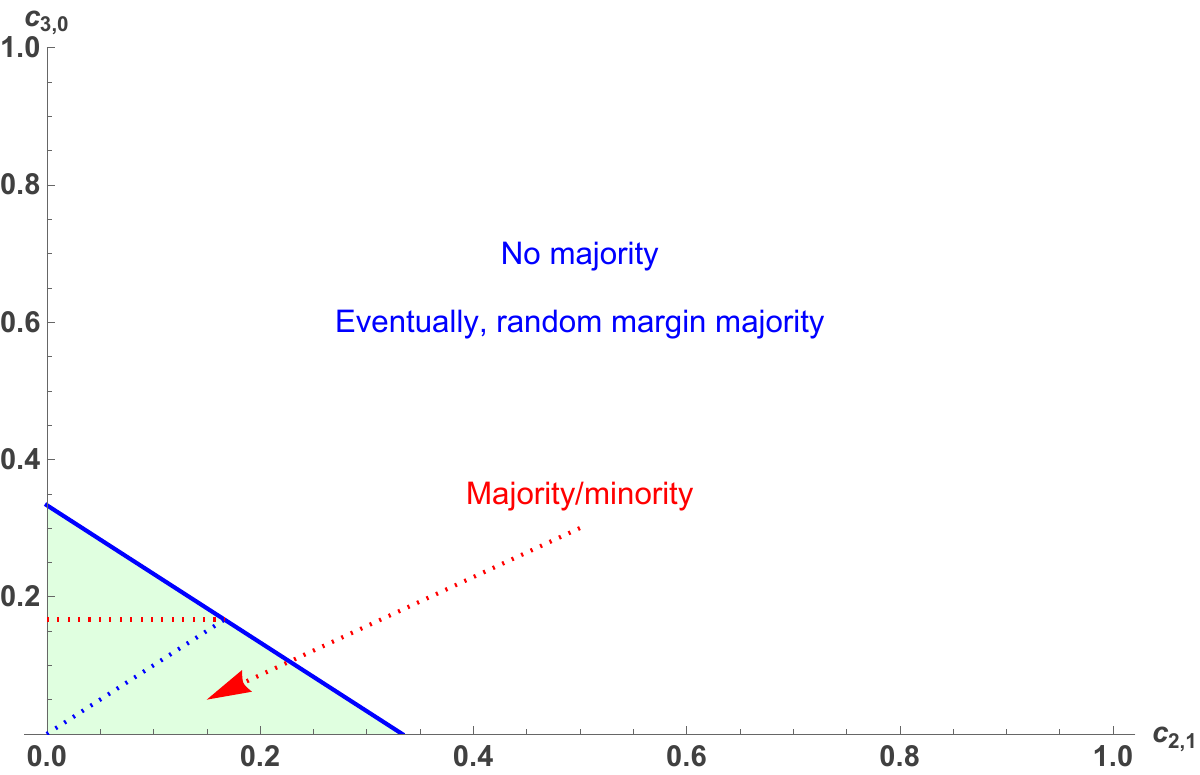}
\includegraphics[width=0.48\textwidth]{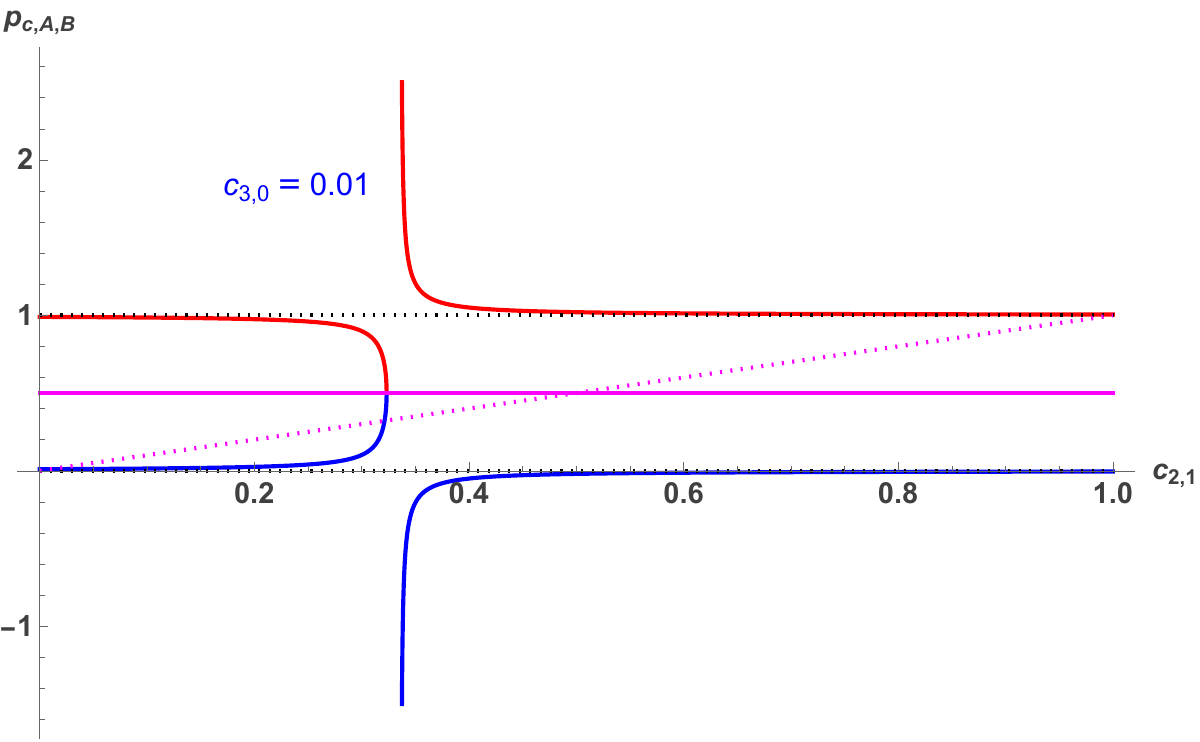}\\
\includegraphics[width=0.48\textwidth]{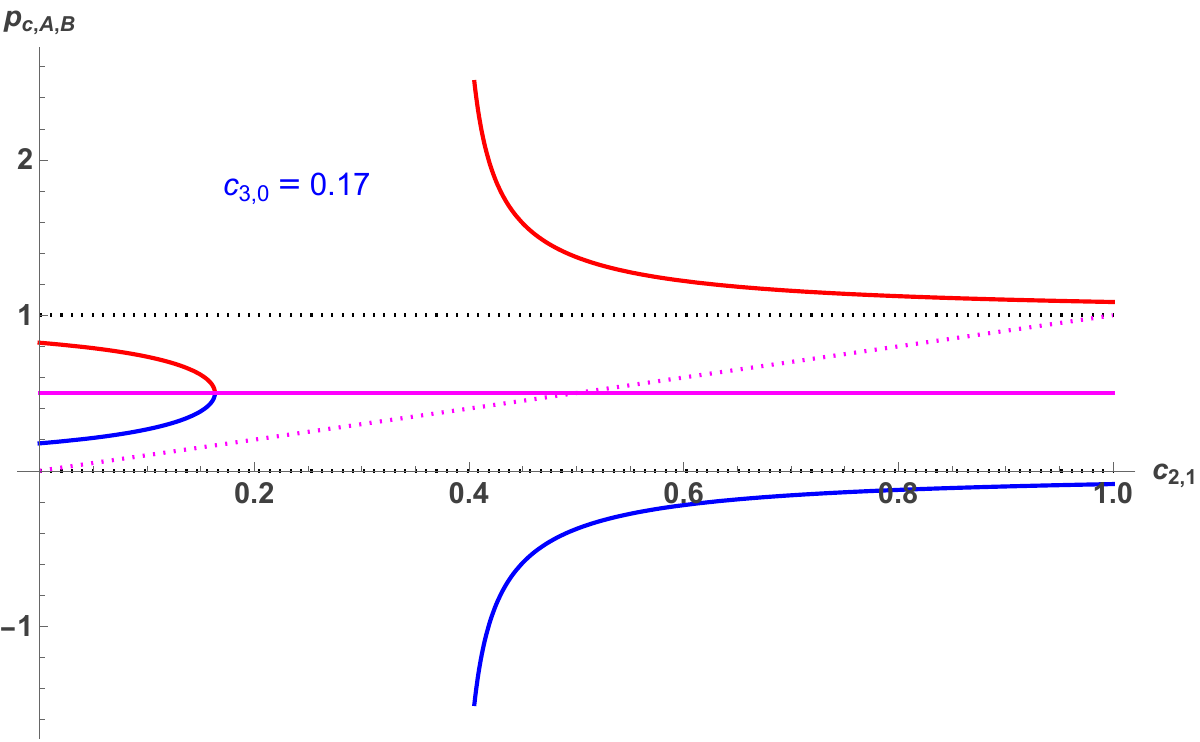}
\includegraphics[width=0.48\textwidth]{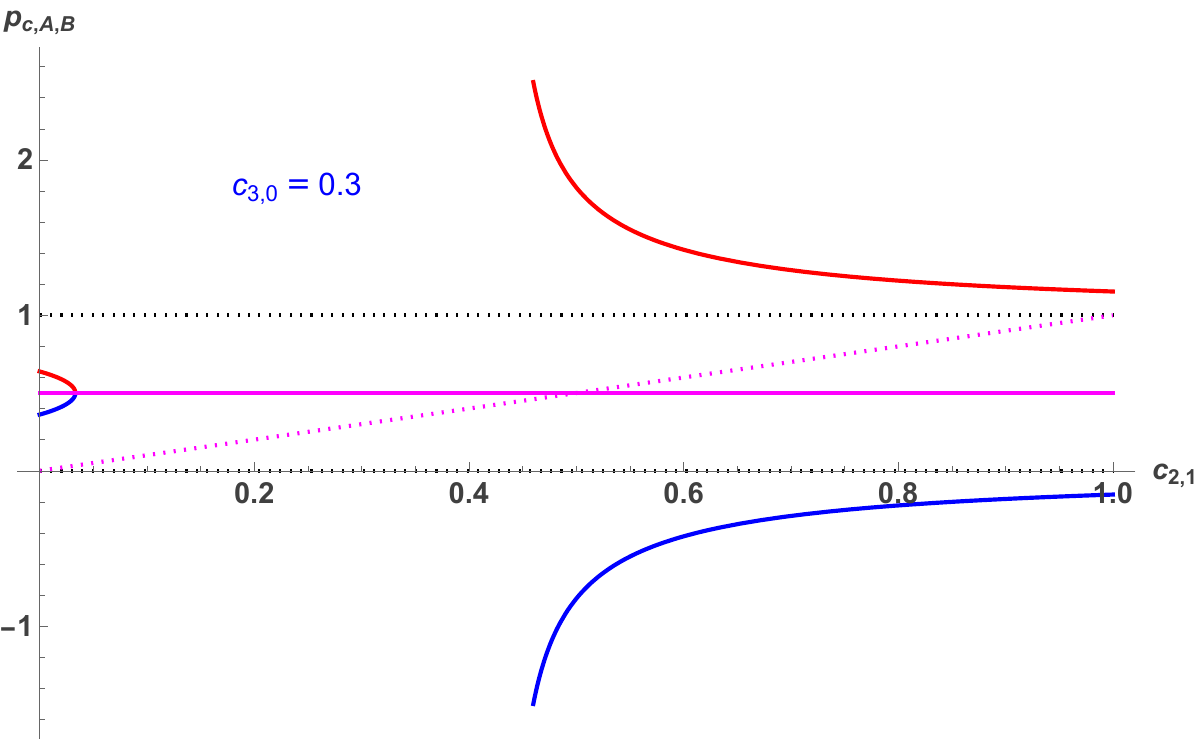}\\
\includegraphics[width=0.48\textwidth]{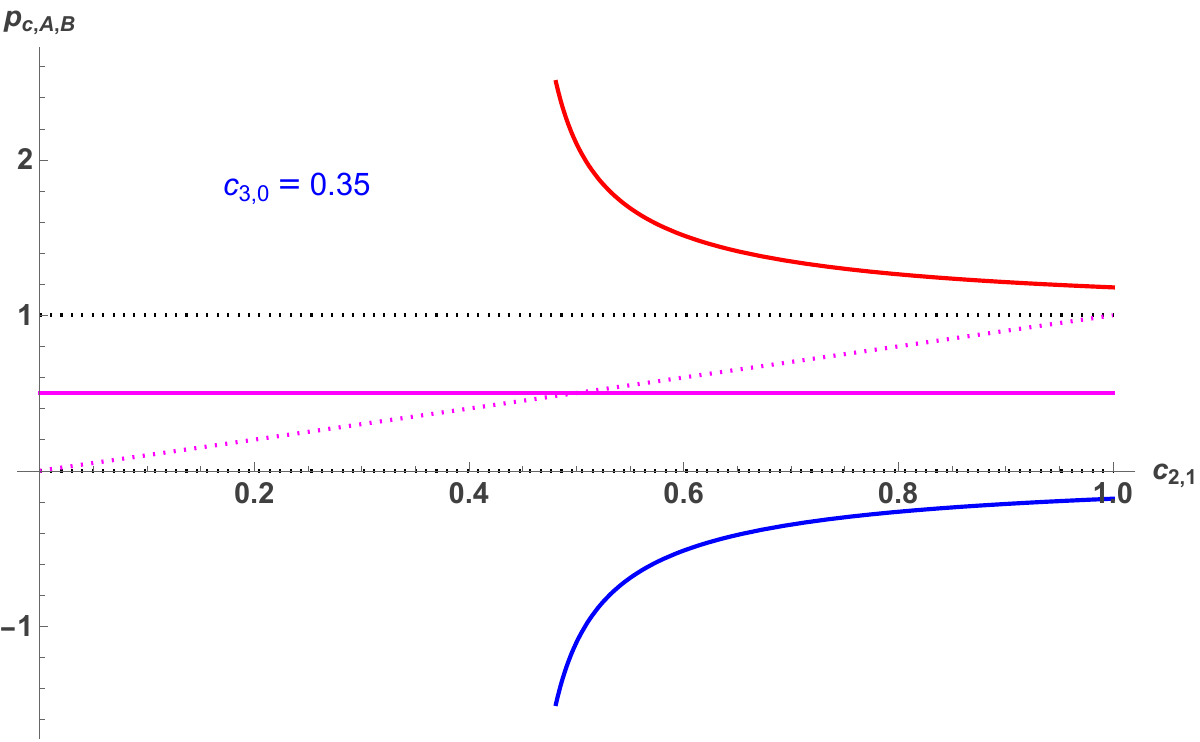}
\includegraphics[width=0.48\textwidth]{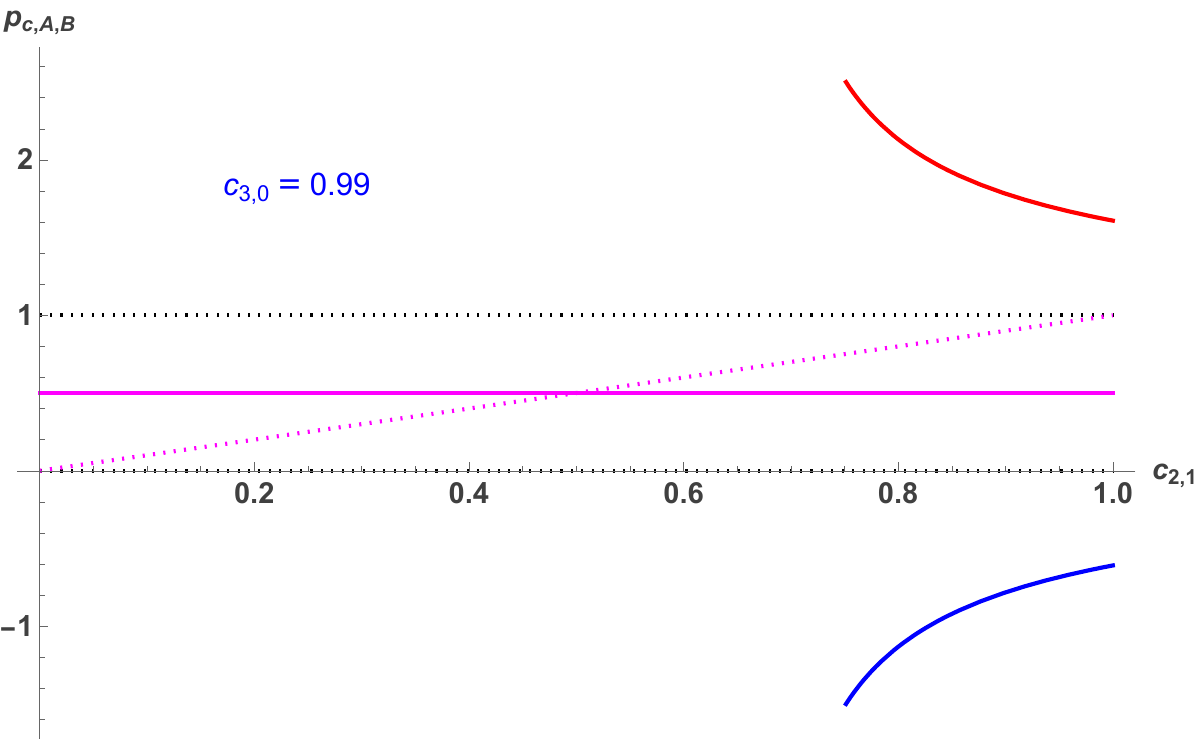}
\caption{Various main stages of the variation $p_{A}$  in red and $p_{B}$ in blue as a function of $c_{2,1}$ for the series of given values $c_{3,0} = 0, 0.01, 0.17, 0.3, 0.35, 0.99$. The fixed point $p_c=\frac{1}{2}$ is also shown in magenta. Only values between zero and one included are valid.}
\label{r1-6}
\end{figure}

Figure\ref{r1-6} exhibits the main stages of the variation $p_{A,B}$  as a function of $c_{2,1}$ for the series of values $c_{3,0} = 0, 0.01, 0.17, 0.3, 0.35, 0.99$. Only values $0 \leq p_{A,B} \leq 1$ are valid.

\subsection{Stability of the fixed point $p_c=\frac{1}{2}$}

Figure\ref{r0} shows a landscape of opinion dynamics obtained from a study of the domain of existence of $p_{A,B}$ combined with a geometric coherence principle. With three valid fixed points two must be attractors and the one in between them a tipping point. With only one valid fixed point, the fixed point must be an attractor. 

A complementary and richer approach involving the stability of the fixed point $p_c=\frac{1}{2}$ is of interest. When $p_c$ is unstable, it is a tipping point and then two attractors must exist around it. When $p_c$ is stable, it is a single attractor. Indeed, the stability of $p_c$ is a function of the derivative of $p_{n+1}$ with respect to $p_n$ taken at the value $p_c$, which yields,
\begin{eqnarray}
\label{d3} 
d_3 &=& \frac{d p_{n+1}}{d p_n}\Big|_{p_c}  ,\nonumber \\ 
&=& \frac{3}{2}(1-c_{3,0}-c_{2,1}) .
\end{eqnarray}

For $0<d_3<1$ the fixed point $p_c$ is an attractor, which happens in the area defined by $c_{3,0}+c_{2,1}>\frac{1}{3}$ recovering the same frontier obtained above and shown in Figure\ref{r0}.

\begin{figure}[h]
\centering
\includegraphics[width=1\textwidth]{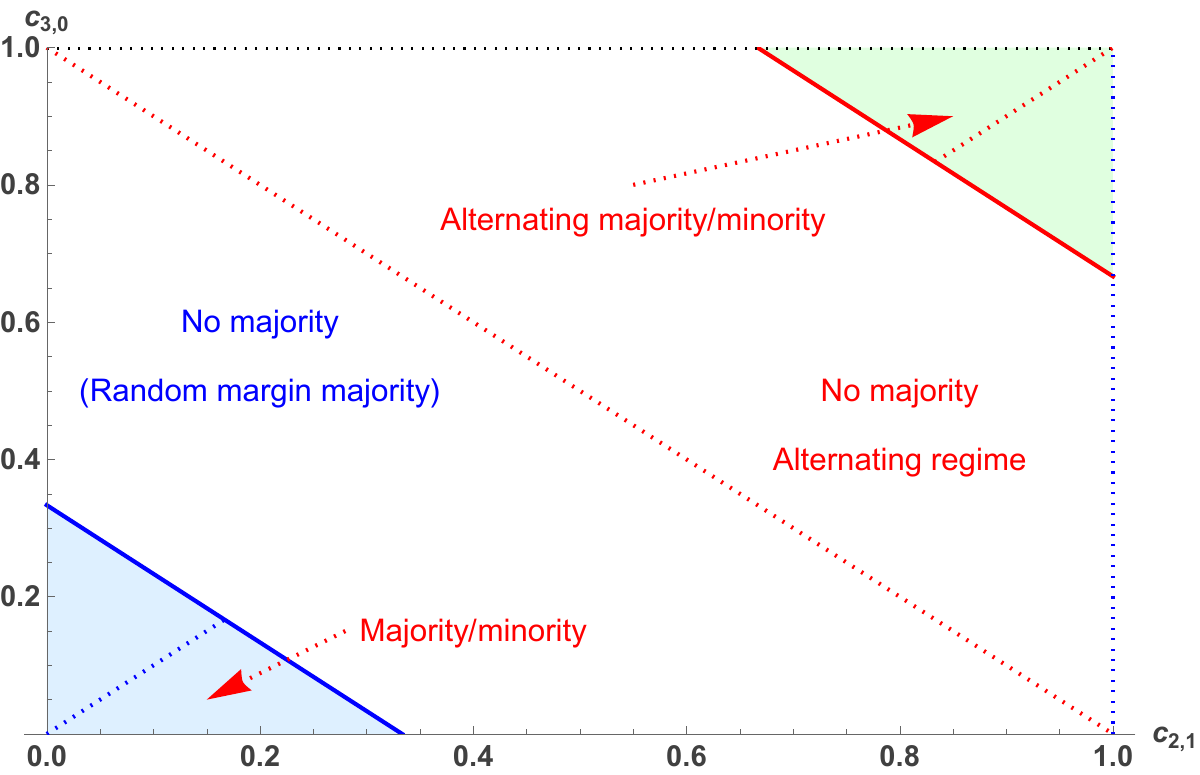}
\caption{The complete landscape of the dynamics including alternating regimes. Four domains are shown as a function of  $c_{3,0}$ and $c_{2,1}$. The domains are separated by the lines $c_{3,0} = \frac{1}{3} (1 - 3  c_{2,1})$ and $c_{3,0} = \frac{1}{3} (5 - 3  c_{2,1})$. Inside the lower left colored area, $p_{A,B}$ exist and are attractors with $p_c$ being a tipping point. Inside the upper right colored area, $p_{\bar{B},\bar{A}}$ exist and are alternating attractors with $p_c$ being an alternating tipping point. Outside the colored areas in the white area only the fixed point $p_c=\frac{1}{2}$ exists and is the single attractor of the dynamics.}
\label{t1}
\end{figure}

Dealing with $d_3$ allows extending the exploration of the opinion landscape by considering the case $-1<d_3<0$, for which $p_c$ is still an attractor but now associated with an alternating dynamics since $d_3<0$. This condition is satisfied within the domain delimited by the condition $c_{3,0}+c_{2,1}<\frac{5}{3}  \Leftrightarrow d_3>-1$ as shown in Figure\ref{t1}.

However, beyond this line where $c_{3,0}+c_{2,1}<\frac{5}{3}$, $p_c$ turns to a tipping point despite the fact that no attractors exist. The apparent contradiction is resolved noticing that $d_3<0$ makes $p_c$ an alternating tipping point. This fact hints at the existence of alternating attractors as discovered recently in the original contrarian model \cite{thwar}. 

Setting $c_{3,0}=c_{2,1}$ recovers the associated values with $c=\frac{1}{6}$ and $c=\frac{5}{6}$ for the two frontiers between the various domains of the opinion landscape.

To prove such a scenario and identify the possible alternating attractors I could solve the Equation $p_{n+2}=p_n$, which a polynomial of degree 9. Yet, solving $p_{n+1}=1-p_n$ yields the same results with a polynomial of degree 3, which yields,

\begin{equation}
p_{\bar{B},\bar{A}}=\frac{1}{2} \Biggr[ 1 \pm \sqrt{\frac{-5 + 3  c_{2,1} + 3  c_{3,0}}{-1 + 3  c_{2,1} -  c_{3,0}}} \Biggl] ,
\label{rrbar3} 
\end{equation}
which exist only in the combined range $\frac{2}{3} \leq c_{2,1}$ and $\frac{1}{3} (5 - 3  c_{2,1}) \leq c_{3,0} \leq 1$. Adding those alternating dynamics to the landscape shown in Figure\ref{r0} leads to the complete landscape exhibited in Figure\ref{t1}.

\begin{figure}[h]
\centering
\includegraphics[width=0.48\textwidth]{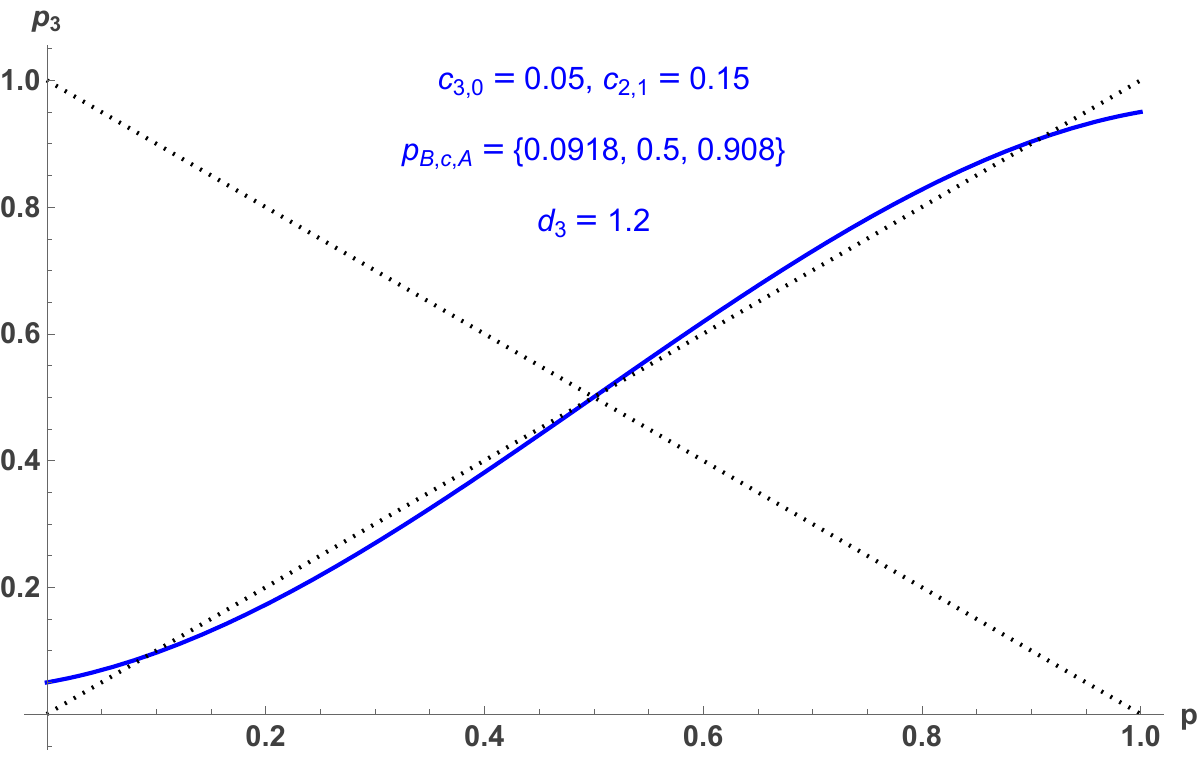}
\includegraphics[width=0.48\textwidth]{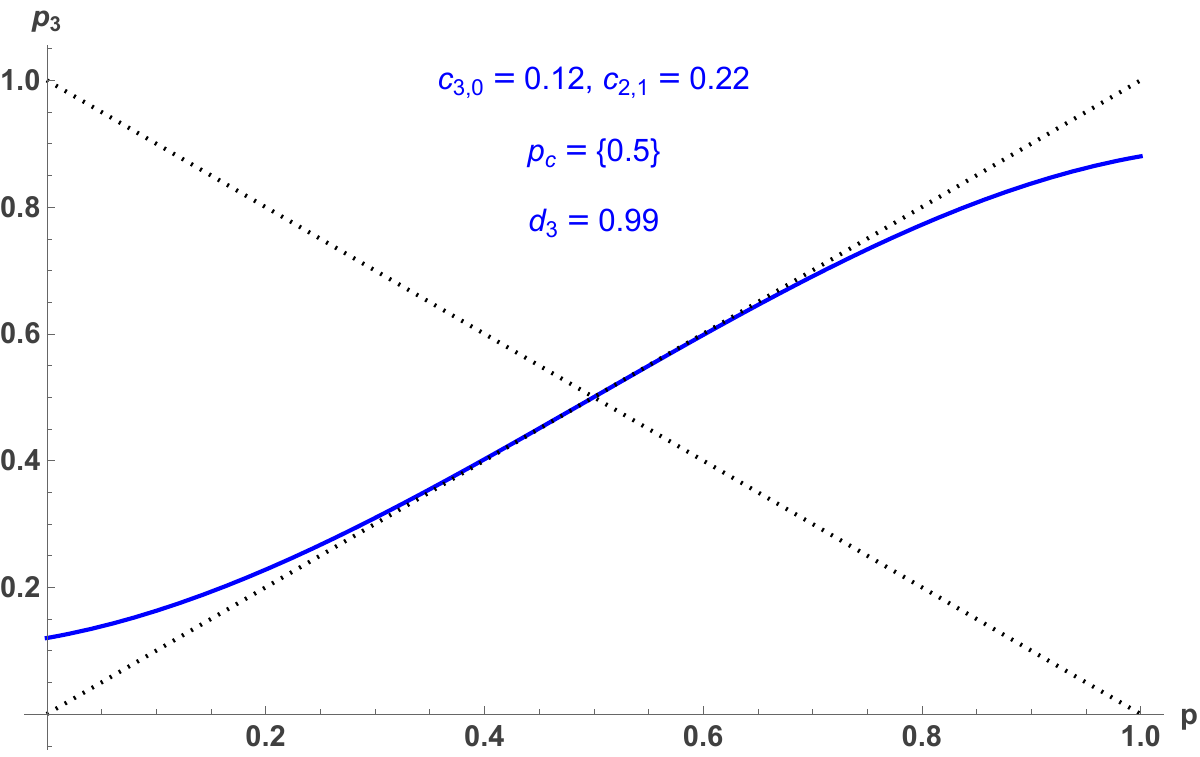}\\
\includegraphics[width=0.48\textwidth]{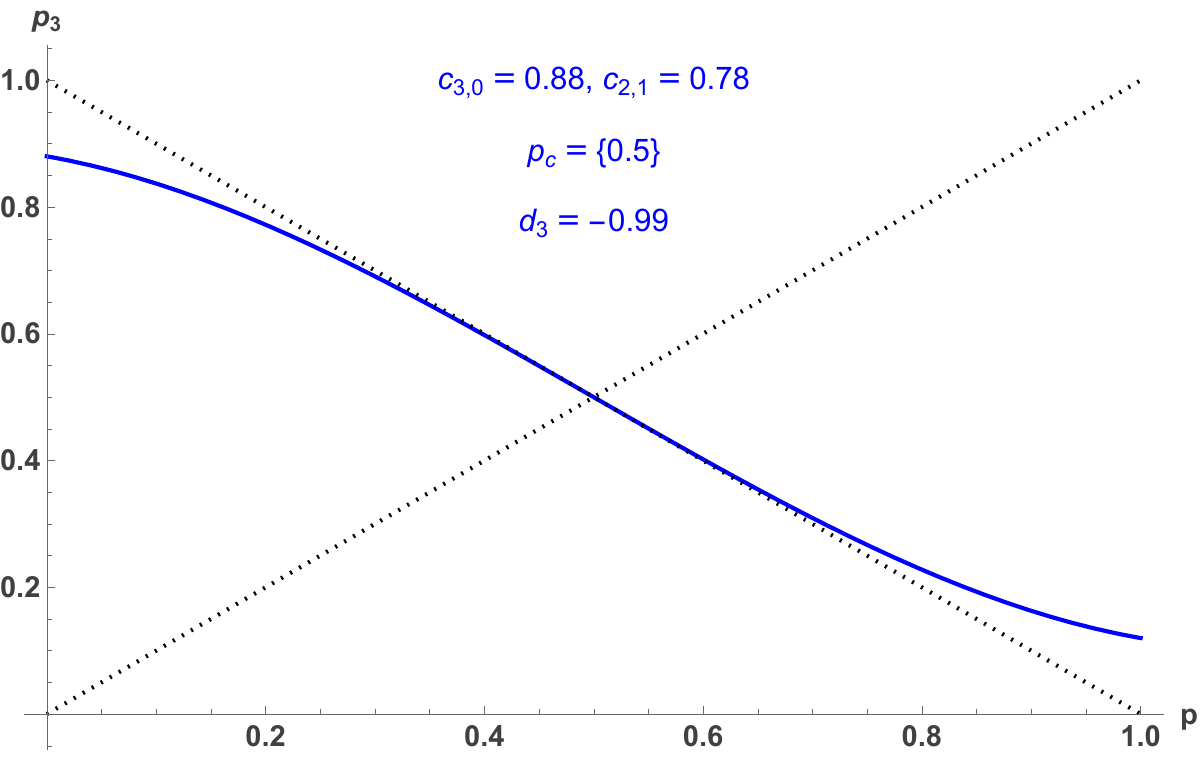}
\includegraphics[width=0.48\textwidth]{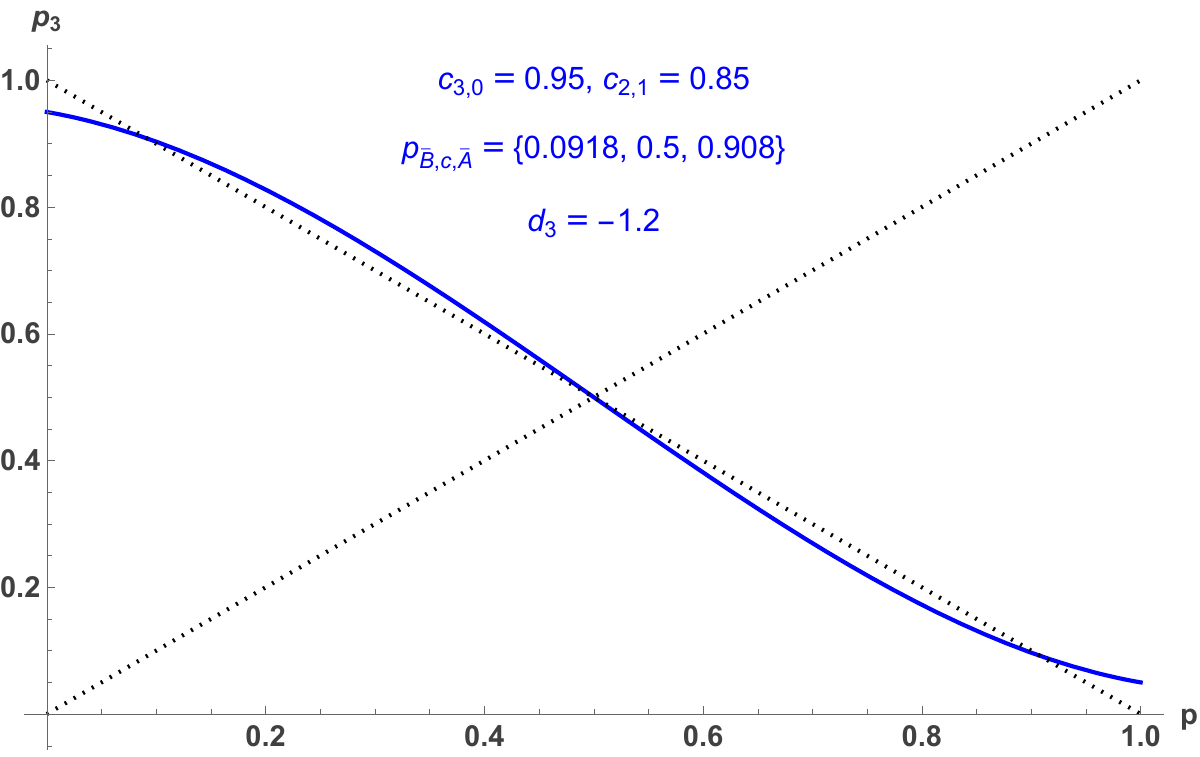}
\caption{ Four different regimes produced by the update Equation (\ref{pcc3}) with a tipping point coupled to two attractors, a single attractor, an alternating single attractor and an alternating tipping point coupled to two alternating attractors, for respectively $(c_{3,0}=0.05, c_{2,1}=0.15)$,  $(c_{3,0}=0.12, c_{2,1}=0.22)$,  $(c_{3,0}=0.88, c_{2,1}=0.78)$,  $(c_{3,0}=0.95, c_{2,1}=0.85)$.}
\label{t2-5}
\end{figure}

Figure\ref{t2-5} shows the associated four different regimes produced by the update Equation (\ref{pcc3}) with a tipping point coupled to two attractors, a single attractor, an alternating single attractor and an alternating tipping point coupled to two alternating attractors, for respectively $(c_{3,0}=0.05, c_{2,1}=0.15)$,  $(c_{3,0}=0.12, c_{2,1}=0.22)$,  $(c_{3,0}=0.88, c_{2,1}=0.78)$,  $(c_{3,0}=0.95, c_{2,1}=0.85)$.

\begin{figure}[h]
\vspace{-3cm}
\centering
\includegraphics[width=0.48\textwidth]{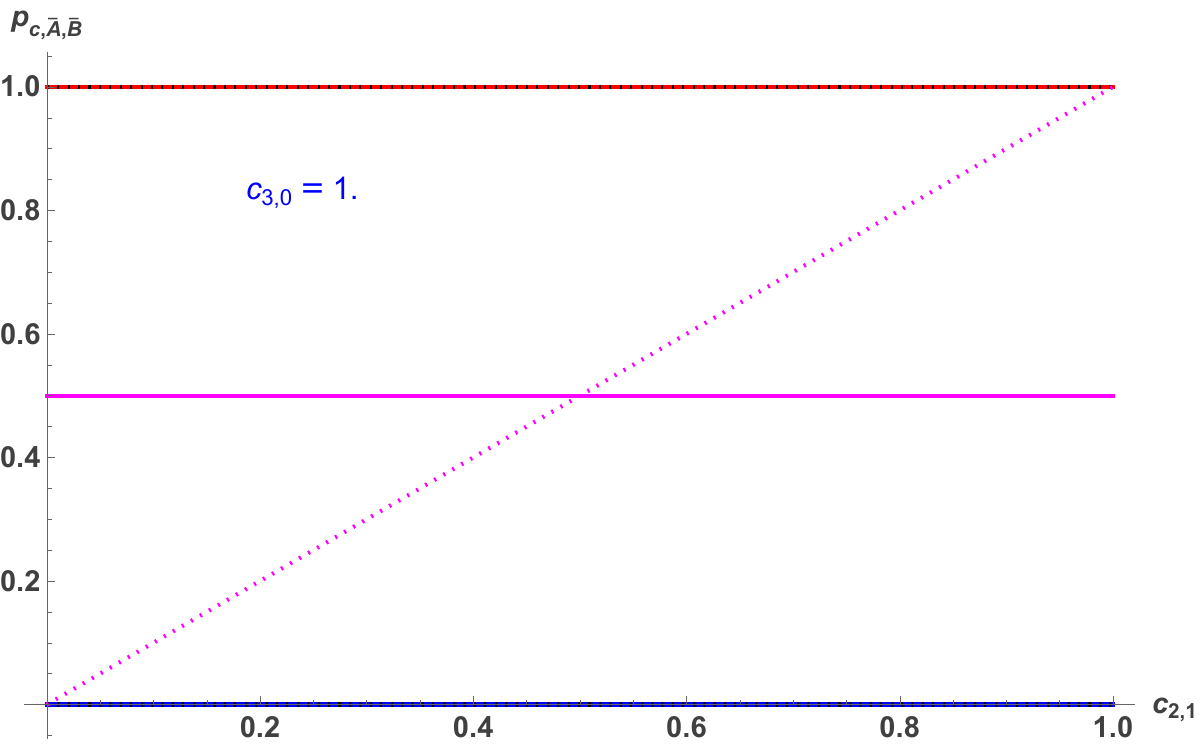}
\includegraphics[width=0.48\textwidth]{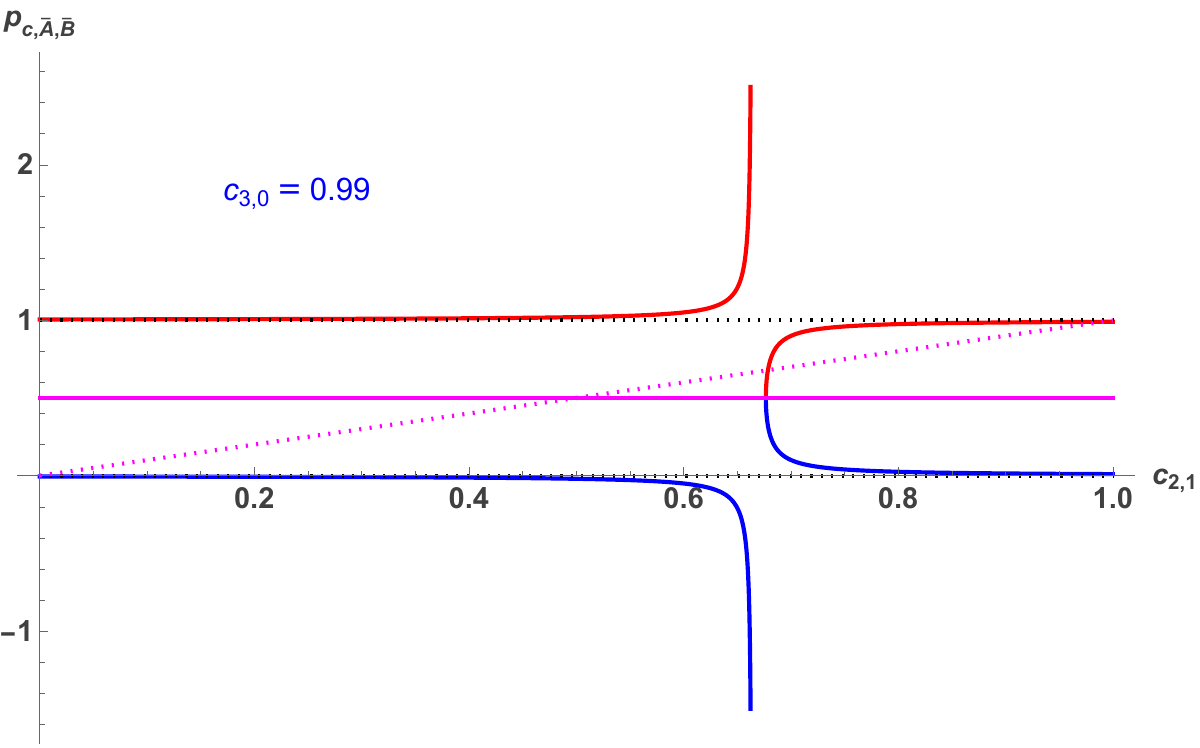}\\
\includegraphics[width=0.48\textwidth]{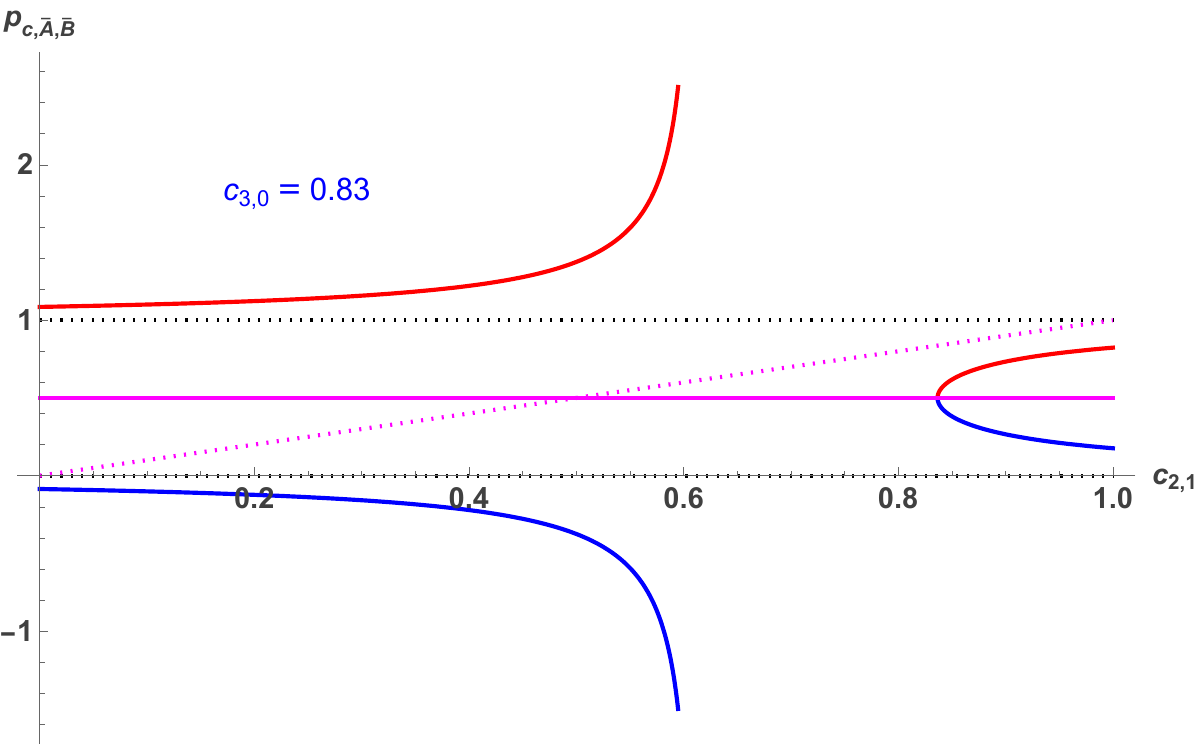}
\includegraphics[width=0.48\textwidth]{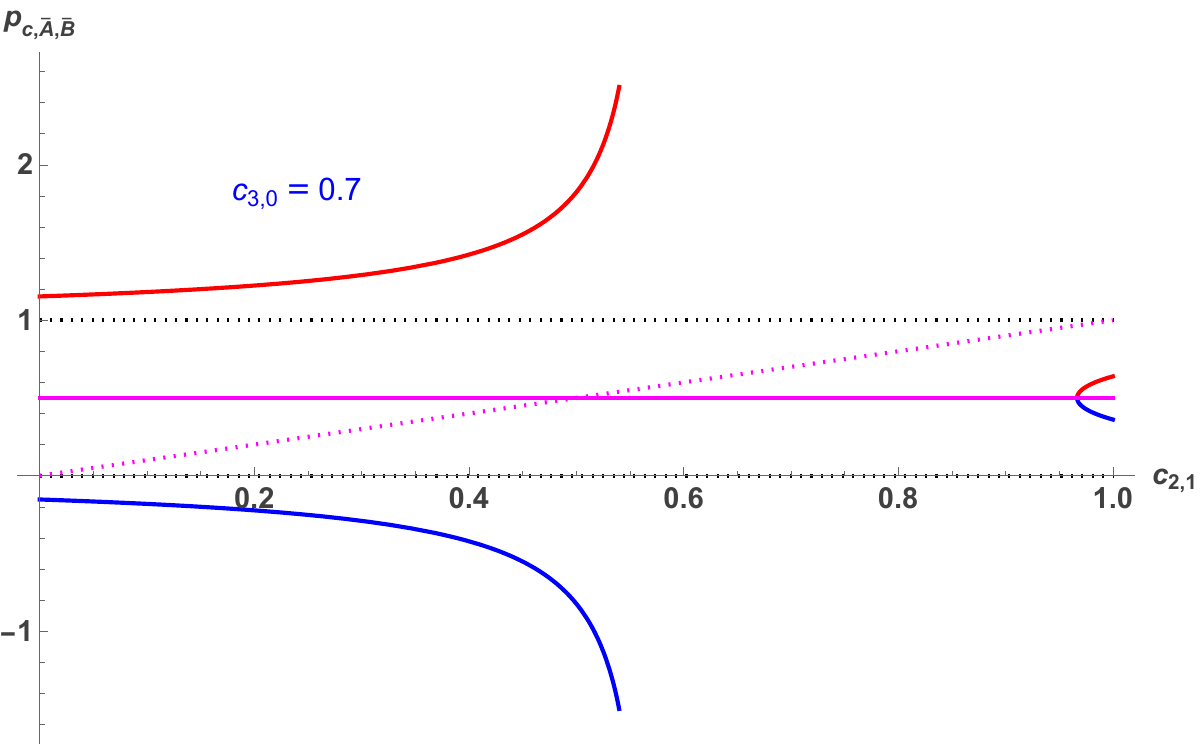}\\
\includegraphics[width=0.48\textwidth]{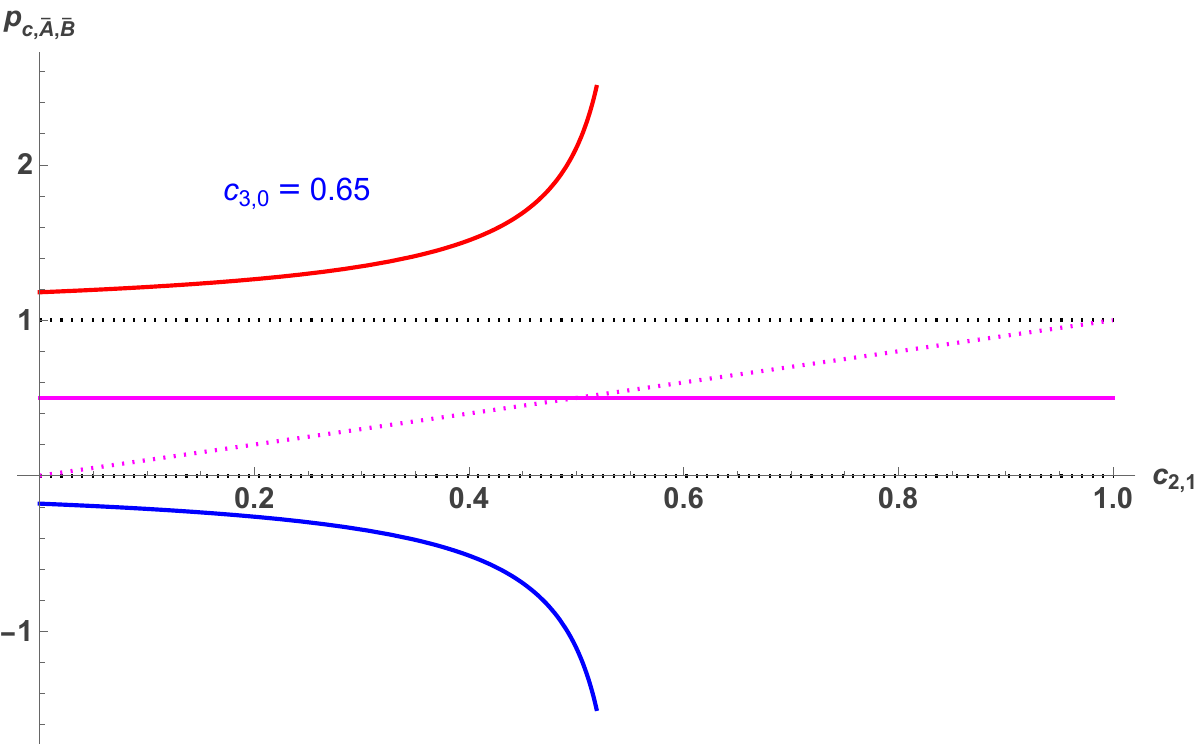}
\includegraphics[width=0.48\textwidth]{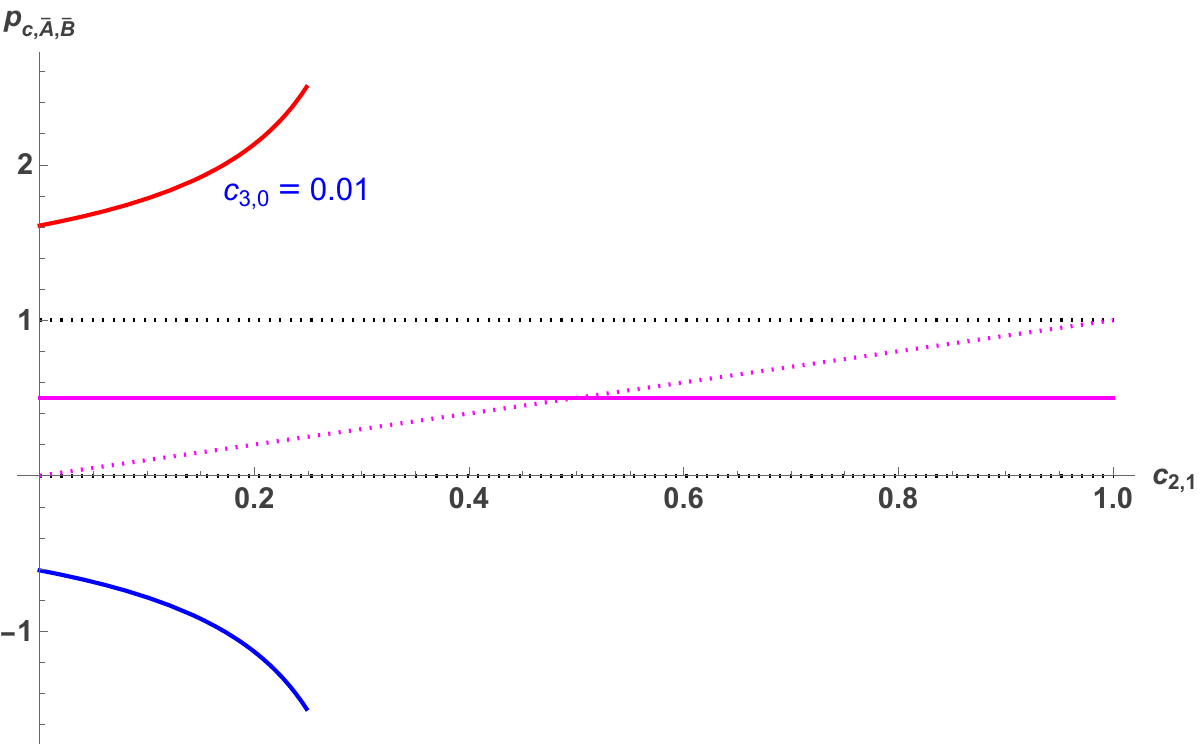}
\caption{Various main stages of the variation $p_{\bar{A}}$ in red and $p_{\bar{B}}$ in blue as a function of $c_{2,1}$ for the series of given values $c_{3,0} = 0, 0.01, 0.17, 0.3, 0.35, 0.99$. The fixed point $p_c=\frac{1}{2}$ is also shown in magenta. Only values between zero and one included are valid.}
\label{u1-6}
\end{figure}

Main stages of the variation $p_{\bar{A}}$ in red and $p_{\bar{B}}$ in blue as a function of $c_{2,1}$ for the series of given values $c_{3,0} = 0, 0.01, 0.17, 0.3, 0.35, 0.99$ are shown in Figure\ref{u1-6}. The fixed point $p_c=\frac{1}{2}$ is also shown in magenta. Only values between zero and one included are valid.

\section{Conclusion}

I have built analytically  the complete two-dimensional landscape of the dynamics of opinion as a function two independent proportions of contrarians, whose respective activations depend on the actual majority-minority ratio in the local discussion group.  Within the Galam Majority Model with group size 3, the two parameters are $c_{3,0}$ and $c_{2,1}$ symmetrically for A and B. In case of unanimity in the group, agents are conformists with probability $(1-c_{3,0})$ and contrarians with probability $c_{3,0}$. In case of two against one, the probabilities are respectively $(1-c_{2,1})$ and $c_{2,1}$. 

In the original model with a proportion $c$ of uniform contrarians the landscape is restricted to one line as a function of $c$. One unique critical value $c=\frac{1}{6}$ separates two regions qualitatively different. In the first one for $0\leq c < \frac{1}{6}$, the dynamics yield a clear cut winner of the public debate ending at an attractor characterized by a stable coexistence of a majority and a minority. In contrast, for $\frac{1}{6} \leq c \leq \frac{5}{6}$, any initial conditions ends up at a single attractor located at fifty percent. While fifty percent means no winner, any real situation will yield a winner determined by random small errors that are normally insignificant. For  $\frac{5}{6} < c\leq 1$ a regime of alternating majority/minority is obtained.

Extending the scheme to a two-dimensional landscape opens a much richer spectrum to intervene on the actual dynamics to modify their ending, either going from majority/minority to fifty/fifty or the reverse. The previous regimes are conserved, but the condition to ensure  majority/minority outcome is now $c_{3,0} < \frac{1}{3} (1 - 3  c_{2,1})$, that can be implemented only for $c_{2,1} < \frac{1}{3}$. Setting $c_{3,0} =c_{2,1} =c$ recovers the condition  $c=\frac{1}{6}$.

Depending on initial supports of each competing opinions, two opposite strategies can be outlined following above findings. For the one having an initial majority of support, the goal is to reduce at maximum the proportions of contrarians to ensure a final victory. In contrast, for the opinion with a minority initial support, the goal is to optimize the proportions of contrarians to reach the area where the dynamics is driven by a single attractor located at fifty percent. There, the outcome of the dynamics becomes random with equal probability to win for both opinions. The a priori losing opinion moves up to a fifty percent chance to win while the a priori winning opinion moves down to a fifty percent chance to win. 


\end{document}